\documentclass[twocolumn]{aastex62}

\usepackage{IEEEtrantools} % for stacked equations
\usepackage{amsmath}

\newcommand{\teff}{\ensuremath{T_{\textrm{eff}}\:}}
\newcommand{\hip}{\ensuremath{\emph{Hipparcos}}\:}

\newcommand{\imag}{\ensuremath{I_{\rm mag}\:}}

\newcommand{\lsol}{\ensuremath{L_{\odot}\:}}

\newcommand{\nstars}{\mbox{25,000}}

\shorttitle{Asteroseismic Target List for TESS}  % running head information
\shortauthors{Schofield et al.}

\begin{document}

\title{THE ASTEROSEISMIC TARGET LIST (ATL) FOR SOLAR-LIKE OSCILLATORS OBSERVED IN 2-MINUTE CADENCE WITH THE TRANSITING EXOPLANET SURVEY SATELLITE (TESS)}

\author[0000-0002-5742-0247]{Mathew Schofield}
\affil{School of Physics and Astronomy, University of Birmingham, Birmingham B15 2TT, UK}
\affiliation{Stellar Astrophysics Centre (SAC), Department of Physics and Astronomy, Aarhus University, Ny Munkegade 120, DK-8000 Aarhus C, Denmark}

\author[0000-0002-5714-8618]{William J. Chaplin}
\email{w.j.chaplin@bham.ac.uk}
\affil{School of Physics and Astronomy, University of Birmingham, Birmingham B15 2TT, UK}
\affiliation{Stellar Astrophysics Centre (SAC), Department of Physics and Astronomy, Aarhus University, Ny Munkegade 120, DK-8000 Aarhus C, Denmark}

\author[0000-0001-8832-4488]{Daniel Huber}
\affiliation{Institute for Astronomy, University of Hawai`i, 2680 Woodlawn Drive, Honolulu, HI 96822, USA}

\author{Tiago L. Campante}
\affiliation{Instituto de Astrof\'{\i}sica e Ci\^{e}ncias do Espa\c{c}o, Universidade do Porto, Rua das Estrelas, PT4150-762 Porto, Portugal}
\affiliation{Departamento de F\'{\i}sica e Astronomia, Faculdade de Ci\^{e}ncias da Universidade do Porto, Rua do Campo Alegre, s/n, PT4169-007 Porto, Portugal}
\affil{School of Physics and Astronomy, University of Birmingham, Birmingham B15 2TT, UK}

\author{Guy R. Davies}
\affil{School of Physics and Astronomy, University of Birmingham, Birmingham B15 2TT, UK}
\affiliation{Stellar Astrophysics Centre (SAC), Department of Physics and Astronomy, Aarhus University, Ny Munkegade 120, DK-8000 Aarhus C, Denmark}

\author{Andrea Miglio}
\affil{School of Physics and Astronomy, University of Birmingham, Birmingham B15 2TT, UK}
\affiliation{Stellar Astrophysics Centre (SAC), Department of Physics and Astronomy, Aarhus University, Ny Munkegade 120, DK-8000 Aarhus C, Denmark}

\author[0000-0002-4773-1017]{Warrick H. Ball}
\affil{School of Physics and Astronomy, University of Birmingham, Birmingham B15 2TT, UK}
\affiliation{Stellar Astrophysics Centre (SAC), Department of Physics and Astronomy, Aarhus University, Ny Munkegade 120, DK-8000 Aarhus C, Denmark}

\author{Thierry Appourchaux}
\affiliation{Universit\'e Paris-Sud, Institut d’Astrophysique Spatiale, UMR 8617, CNRS, B\^atiment 121, F-91405 Orsay Cedex, France}

\author[0000-0002-6163-3472]{Sarbani Basu}
\affiliation{Department of Astronomy, Yale University, PO Box 208101, New Haven, CT 06520-8101, USA}

\author{Timothy R. Bedding}
\affiliation{Sydney Institute for Astronomy (SIfA), School of Physics, 2006 University of Sydney, Australia}
\affiliation{Stellar Astrophysics Centre (SAC), Department of Physics and Astronomy, Aarhus University, Ny Munkegade 120, DK-8000 Aarhus C, Denmark}

\author{J{\o}rgen Christensen-Dalsgaard}
\affiliation{Stellar Astrophysics Centre (SAC), Department of Physics and Astronomy, Aarhus University, Ny Munkegade 120, DK-8000 Aarhus C, Denmark}

\author{Orlagh Creevey}
\affiliation{Universit\'e Cô\^oe d'Azur, Observatoire de la C\^ote d'Azur, CNRS, Laboratoire Lagrange, Bd de l'Observatoire, CS 34229, 06304, Nice Cedex 4, France}

\author{Rafael A. Garc\'ia}
\affiliation{Laboratoire AIM, CEA/DSM – CNRS - Univ. Paris \\
Diderot – IRFU/SAp, Centre de Saclay \\
91191 Gif-sur-Yvette Cedex, France}

\author{Rasmus Handberg}
\affiliation{Stellar Astrophysics Centre (SAC), Department of Physics and Astronomy, Aarhus University, Ny Munkegade 120, DK-8000 Aarhus C, Denmark}

\author{Steven D. Kawaler}
\affiliation{Department of Physics and Astronomy, Iowa State University, Ames, IA 50011, USA}

\author{Hans Kjeldsen}
\affiliation{Stellar Astrophysics Centre (SAC), Department of Physics and Astronomy, Aarhus University, Ny Munkegade 120, DK-8000 Aarhus C, Denmark}

\author{David W. Latham}
\affiliation{Harvard-Smithsonian Center for Astrophysics, 60 Garden Street, Cambridge, MA 02138, USA}

\author{Mikkel N. Lund}
\affiliation{Stellar Astrophysics Centre (SAC), Department of Physics and Astronomy, Aarhus University, Ny Munkegade 120, DK-8000 Aarhus C, Denmark}

\author{Travis S. Metcalfe}
\affiliation{Space Science Institute, 4750 Walnut Street, Suite 205, Boulder, CO 80301, USA}
\affiliation{Max-Planck-Institut f\"ur Sonnensystemforschung, Justus-von-Liebig-Weg 3, 37077, G\"ottingen, Germany}

\author{George R. Ricker}
\affiliation{MIT Kavli Institute for Astrophysics and Space Research, 70 Vassar St., Cambridge, MA 02139, USA}

\author{Aldo Serenelli}
\affiliation{Institute of Space Sciences (ICE, CSIC) Campus UAB, Carrer de Can Magrans s/n, 08193 Barcelona, Spain}
\affiliation{Institut d’Estudis Espacials de Catalunya (IEEC), C/ Gran Capit\`a, 2-4, 08034 Barcelona, Spain}

\author{Victor Silva Aguirre}
\affiliation{Stellar Astrophysics Centre (SAC), Department of Physics and Astronomy, Aarhus University, Ny Munkegade 120, DK-8000 Aarhus C, Denmark}

\author{Dennis Stello}
\affiliation{Sydney Institute for Astronomy (SIfA), School of Physics, 2006 University of Sydney, Australia}
\affiliation{Stellar Astrophysics Centre (SAC), Department of Physics and Astronomy, Aarhus University, Ny Munkegade 120, DK-8000 Aarhus C, Denmark}

\author{Roland Vanderspek}
\affiliation{MIT Kavli Institute for Astrophysics and Space Research, 70 Vassar St., Cambridge, MA 02139, USA}

\begin{abstract}

We present the target list of solar-type stars to be observed in short-cadence (2-min) for asteroseismology by the NASA \emph{Transiting Exoplanet Survey Satellite} (TESS) during its 2-year nominal survey mission. The solar-like Asteroseismic Target List (ATL) is comprised of bright, cool main-sequence and subgiant stars and forms part of the larger target list of the TESS \emph{Asteroseismic Science Consortium} (TASC). The ATL uses Gaia DR2 and the Extended $\hip$ Compilation (XHIP) to derive fundamental stellar properties, calculate detection probabilities and produce a rank-ordered target list. We provide a detailed description of how the ATL was produced and calculate expected yields for solar-like oscillators based on the nominal photometric performance by TESS. We also provide publicly available source code which can be used to reproduce the ATL, thereby enabling comparisons of asteroseismic results from TESS with predictions from synthetic stellar populations.
\end{abstract}

\section{Introduction}
\label{sect:intro}

NASA's \emph{Transiting Exoplanet Survey Satellite} (TESS) was launched on 2018 April 18 with the main goal to detect small planets orbiting nearby stars using the transit method \citep{ricker_transiting_2014}. Its photometric data will also enable high-fidelity studies of stars, and other astrophysical objects and phenomena (e.g. transients, galaxies, solar-system objects etc.). TESS is observing bright stars, including those visible to the naked eye, opening up a new discovery space to characterize stars several magnitudes brighter than those observed by the NASA \emph{Kepler} Mission. While \emph{Kepler} \citep{borucki_kepler_2010} and its re-purposed follow-on Mission known as K2 \citep{howell_k2_2014} observed stars in only dedicated fields, TESS will survey over 85\,\% of the sky during its 2-year nominal mission\footnote{{\url{https://heasarc.gsfc.nasa.gov/docs/tess/}}}, covering first the southern and then the northern equatorial hemispheres (e.g., see \citealt{huang18}). TESS thus promises to provide a unique census of bright stars in the solar neighborhood. 

TESS will produce Full-Frame Image (FFI) data every 30\,min for the entire field of view, and 2-min (short-cadence) data on a total of approximately 200,000 targets. The short-cadence target list is comprised of several cohorts: high-priority targets for exoplanet transit searches, which form the Candidate Target List (CTL) \citep{Stassun2018}; targets from the TESS Guest Investigator (GI) program\footnote{\url{https://heasarc.gsfc.nasa.gov/docs/tess/proposing-investigations.html}}, the Director’s Discretionary Target (DDT) and out-of-cycle Target of Opportunity (ToO) programs\footnote{\url{https://tess.mit.edu/science/ddt/}}; and targets for asteroseismic studies of stars (e.g.\ \citealt{chaplin_asteroseismology_2013}).

The high-precision, high-cadence, near continuous photometric data that TESS will provide are well suited to asteroseismology. As with \emph{Kepler} \citep{gilliland_kepler_2010}, the international asteroseismology community is coordinating efforts through the TESS \emph{Asteroseismic Science Consortium} (TASC)\footnote{\url{http://tasoc.dk}}. Owing to their short oscillation periods, there are several classes of stars that require short-cadence data for asteroseismology. The most prominent examples are solar-type stars, here defined as cool main-sequence and sub-giant stars which show solar-like oscillations that are stochastically excited and intrinsically damped by near-surface convection. \emph{Kepler} and K2 have provided asteroseismic detections in approximately 700 solar-type stars \citep{ChaplinSci2011, Chaplin2014, Lund2016a}, including about 100 \emph{Kepler} planet hosts \citep{Huber2013,lundkvist16}. The main limitation for the asteroseismic yield of \emph{Kepler}/K2 was the limited number of short-cadence target slots; there were around 500 available at any one time to the mission. That constraint will be eased dramatically for TESS, giving the potential to provide detections in thousands of solar-type stars. In addition to asteroseismic characterizations of already known planet hosts \citep{campante_asteroseismic_2016}, TASC will also provide the TESS Science Team with such data on the bright solar-type hosts around which TESS will discover planets.

TESS will dedicate around 20,000 short-cadence targets to asteroseismology, and it is the responsibility of TASC to provide the target list. In this paper we describe the construction of the prioritized Asteroseismic Target List (ATL) of solar-like oscillators, which forms part of the overall TASC list.  The breakdown of the rest of the paper is as follows. We begin in Section~\ref{sec:philo} by describing the basic philosophy underlying the construction of the ATL. Section~\ref{sec:data} summarizes the input data. In Section~\ref{sec:construction} we discuss in detail the steps followed to produce a prioritized target list. Then in Section~\ref{sec:overview} we provide an overview of the rank-ordered list, including a prediction of the overall asteroseismic yield. We finish in Section~\ref{sec:sum} with a summary overview of the list, including information on how to access both the ATL in electronic form\footnote{\url{https://figshare.com/s/aef960a15cbe6961aead}} and the \texttt{Python} codes used to construct it in a \texttt{Github} repository\footnote{\url{https://github.com/MathewSchofield/ATL_public}}.

\section{Philosophy for the Construction of the ATL}
\label{sec:philo}

Our goal was to produce an all-sky rank-ordered target list based on basic observables from all-sky catalogs and derived quantities which can be easily be duplicated for simulated populations (to facilitate stellar populations studies). The most obvious approach would be to select stars that are expected to show solar-like oscillations (i.e., stars cool enough to have convective envelopes), and then rank by apparent magnitude (either in the TESS bandpass, $T$, or Johnson $I$-band which is a good proxy of the TESS magnitude). However, we must also consider whether solar-like oscillations are likely to be detected in a potential target. This requires a prediction of expected photometric amplitudes of the solar-like oscillations, stellar granulation, and the expected shot and instrumental noise. A simple rank-order approach based on apparent magnitude would significantly compromise the potential yield of asteroseismic detections, and omit targets for which we expect to make asteroseismic detections.

We therefore base the ranking in our list on  predictions of asteroseismic detectability, which were made using the basic methodology developed for and applied successfully to \emph{Kepler} target selection \citep{chaplin_predicting_2011}. While this approach is more complicated it is worth stressing that the asteroseismic predictions use simple analytical formulae, which may be applied straightforwardly to synthetic populations. All codes and data used to produce the target list are publicly available to facilitate reproducibility and the comparison with synthetic stellar populations.

\section{Input Data}
\label{sec:data}

\subsection{Input catalogs}

The ATL is mainly based on targets in Gaia Data Release 2 (DR2)\footnote{\url{http://www.cosmos.esa.int/gaia}\label{ref:gaia}} \citep{gaia18}, supplemented at bright magnitudes by the eXtended $\hip$ Compilation (XHIP) \citep{anderson_xhip:_2012}.
%, available on VizieR as the online data catalogue V/137D. 
The basic set of data used to construct the ATL comprises the astrometric distances, magnitudes in the $I$ and $V$ bands, $(B-V)$ color, and the sky positions. From these input data we may estimate the photometric variability in the TESS bandpass caused by solar-like oscillations, granulation and shot/instrumental noise, as well as the expected duration of the TESS observations. Using these derived quantities, we then calculate the probability of detecting solar-like oscillations.

\subsection{Fundamental Stellar Properties}

Distances for {\it Gaia} DR2 stars were taken from \citet{bailerjones18}  using the median of the posterior calculated using their Milky Way prior. This set of distances was chosen because the Milky Way prior performs better for stars closer than 2\,kpc, where the vast majority of the ATL targets are located. Distances for XHIP stars were derived by inverting the parallax. We added a zeropoint offset of 0.029\,mas to all Gaia DR2 parallaxes \citep{luri18,zinn18}. After this, we discarded all targets in both catalogs which have a fractional parallax uncertainty $\sigma_{\pi} / \pi > 0.5$.
%(having first added the 0.25\,mas systematic parallax correction given in \citet{stassun_evidence_2016} to the DR2 parallaxes \textbf{[Mat: the offset in this paper only applied to Gaia DR1. Did we really do this? It turns out that DR2 has an actual offset (unlike DR1) in the same sense, so if yes we can still argue our way around this. Bill: apologies, this text was from when we were using TGAS. For DR2, we added 0.029mas to every parallax (given in Luri et al. 2018).]}). 
Reddening and extinction in the $V$ and $I$ bands were calculated from the derived distances and sky positions (Galactic coordinates) using the Combined15 dust map from the \textsc{mwdust} Python package \citep{marshall_modelling_2006,green_three-dimensional_2015,drimmel_three-dimensional_2003,bovy_galactic_2016}.

While $I$-band magnitudes are available for XHIP targets, this is not the case for most of the \textit{Gaia} DR2 targets. This is important because the $I$ magnitudes are needed to estimate the shot noise in the TESS bandpass. We therefore used $(B-V)$ colors and apparent $V$ magnitudes to derive the required values. The preferred source for both inputs was the revised {\it Hipparcos} catalog \citep{van_leeuwen_hipparcos_2007}. If those data were unavailable, we used the {\it Tycho-2} catalog \citep{hog_tycho-2_2000}; and failing that, we took values from the AAVSO All-Sky Photometric Survey (APASS; \citealt{henden_aavso_2009}).

The input $(B-V)$ colors were first de-reddened, using the previously calculated $E(B-V)$, and then converted to $(V-I)$ using the polynomials in \citet{caldwell_statistical_1993}. The coefficients of the polynomial depend upon whether the target is classified as a ``giant'' or ``dwarf''. Here, we separated targets using an empirically derived relation in $M_{g}$, the absolute magnitude in the \textit{Gaia} bandpass, and $(B-V)$, using DR2 data, classifying stars with $M_{g} > 6.5 \times (B-V) - 1.8$ as dwarfs, and the rest as giants. Once $(V-I)$ had been calculated for all of the stars, the $I$ magnitudes were estimated from $V$ and $(V-I)$. The derived $I$ magnitudes were then reddened using the previously estimated $A_{I}$ to calculate the TESS noise (see Section \ref{sec:construction}). 

Dereddened $(B-V)$ colors were used to estimate stellar effective temperatures $T_{\rm eff}$, using color-temperature relations of the form\footnote{We adopt relations that do not include any correction for metallicity, since we do not have good/uniform quality estimates of [Fe/H] for all targets under consideration}:
\begin{equation}
{\rm log}(T_{\rm eff}) = a + b(B-V) + c(B-V)^{2} + ... ,
\label{eq:teff}
\end{equation} 
where the best-fitting coefficients were taken from \citet{torres_use_2010}. Luminosities, $L$, were calculated from
 \begin{equation}
 \begin{split}
 \log(L/{\rm L_{\odot}}) = 4.0 + 0.4M_{\rm bol \odot} - 2.0\log\pi - \\
 0.4 \left(V_{\rm mag}-A_V+{\rm BC}_V \right).
 \label{eq:lum}
 \end{split}
 \end{equation}
Note that $V$ magnitudes were first de-reddened using the previously calculated $A_V$, while the bolometric corrections, $BC_{\rm V}$, were taken from \citet{flower_transformations_1996}, as presented in \citet{torres_use_2010}, with $M_{\rm bol,\odot}=4.73 \pm 0.03$ mag. Finally, we estimated radii using the Stefan-Boltzmann law $L \propto R^2 T_{\rm eff}^4$, using $T_{\rm eff,\odot}=5777\,\rm K$.

\subsection{Comparison to Literature Values}

\begin{figure}
	\centering
    \resizebox{\hsize}{!}{\includegraphics{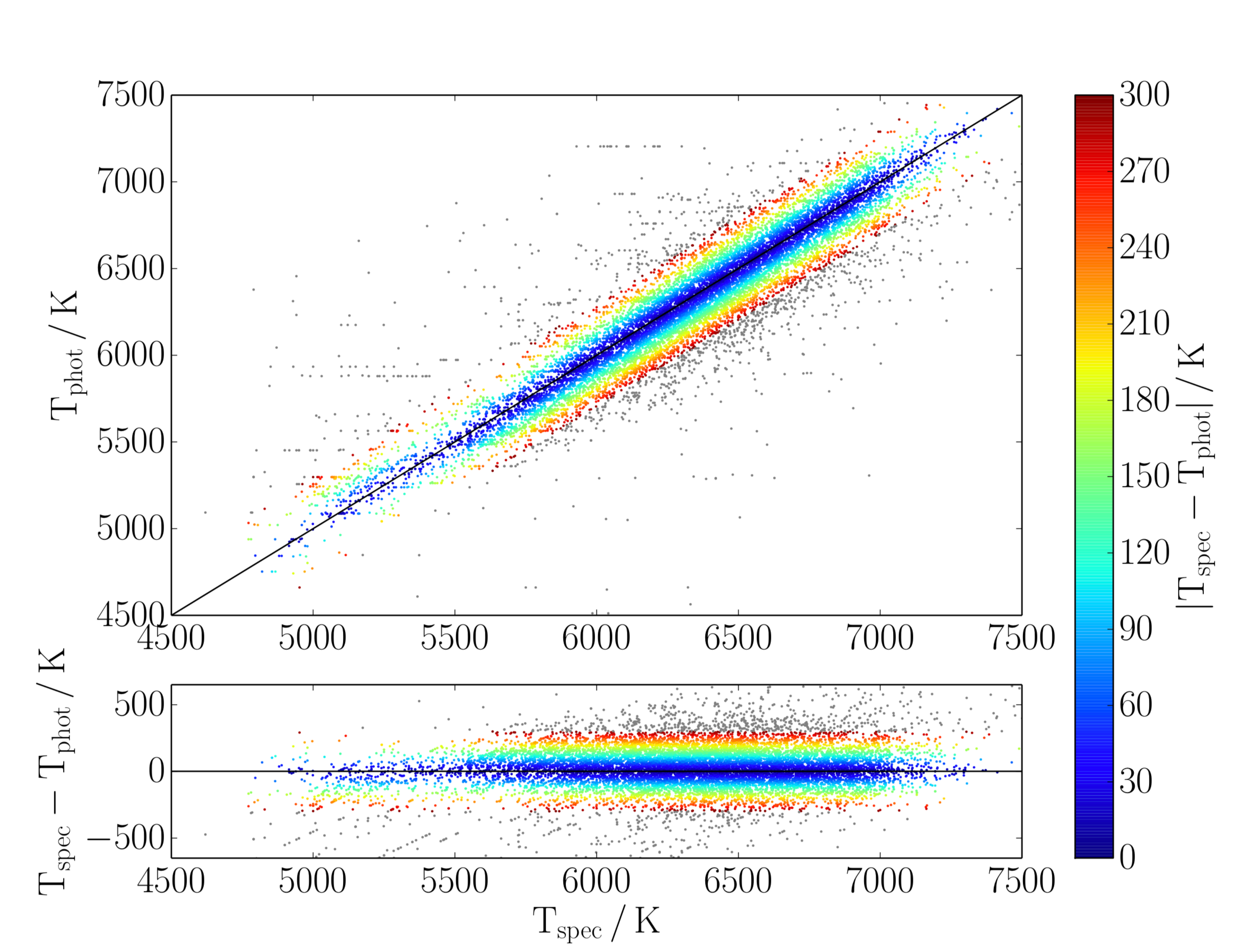}}
	\caption{Comparison between effective temperatures from high-resolution spectroscopy (as listed in the PASTEL catalogue) and the ATL photometric temperatures. The solid line shows the 1:1 relation. The horizontal lines of datapoints exist because the PASTEL catalogue gives several effective temperatures for some stars. The residual median and scatter is 102\,K and 146\,K, respectively.}	
	\label{fig:pastel}
\end{figure}

\begin{figure}
	\centering
    \resizebox{\hsize}{!}{\includegraphics{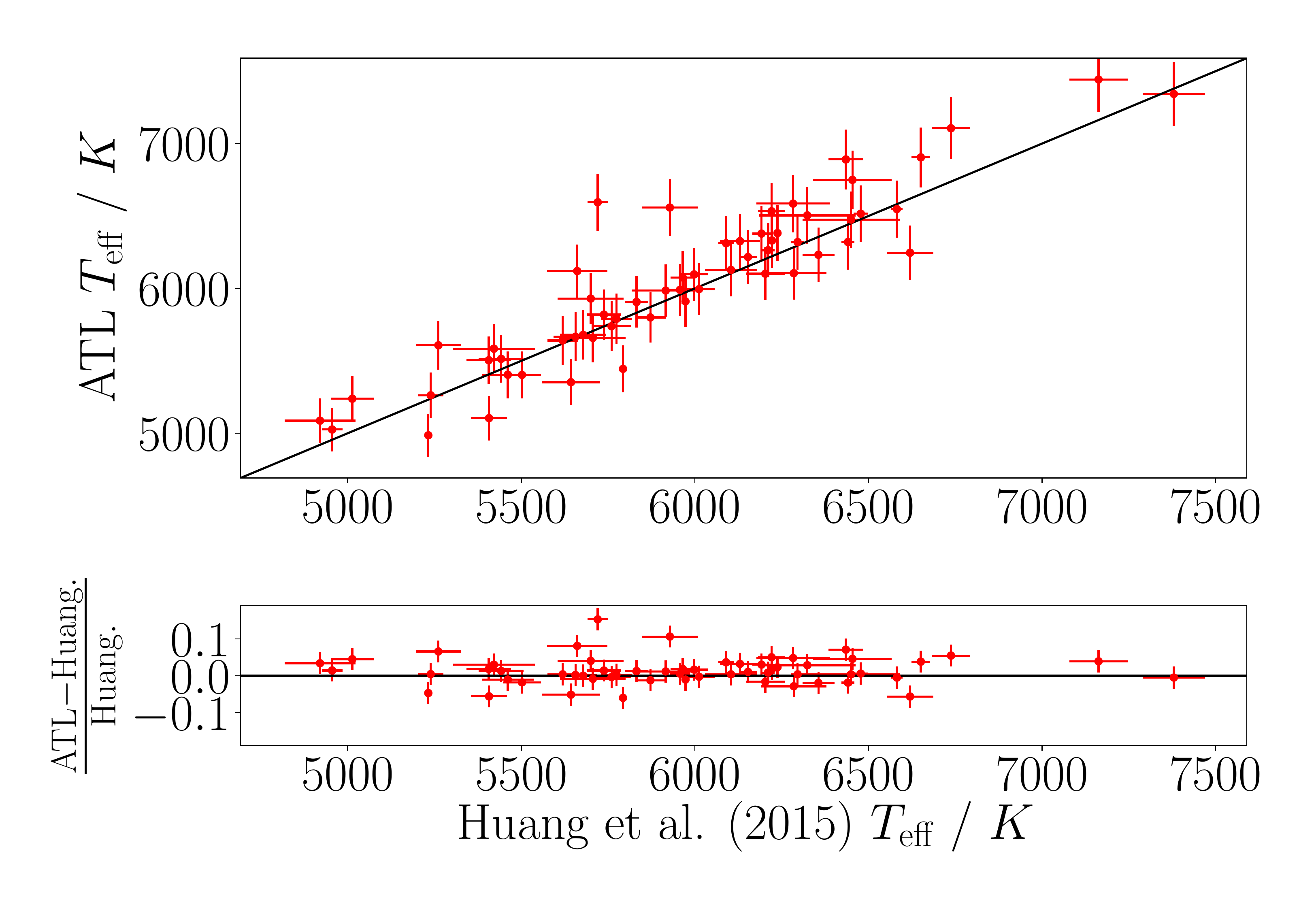}}
	\caption{Comparison between effective temperatures from long-baseline interferometry \citep[as compiled by][]{huang_empirical_2015} and the ATL. The solid line shows the 1:1 relation. The residual median and scatter are 109\,K and 173\,K, respectively.}	
	\label{fig:huang}
\end{figure}

We compared our estimated stellar properties with several literature sources. The PASTEL catalog \citep{soubiran_pastel_2016} includes spectroscopically-determined effective temperatures for over 60,000 stars. Figure~\ref{fig:pastel} compares our derived photometric temperatures with PASTEL for stars that are common to both lists. We observe a good agreement, with a residual median and scatter of 102\,K and 146\,K, respectively. We furthermore compared our temperatures with values listed in \citet{huang_empirical_2015}, which compiled empirical temperatures derived from optical long-baseline interferometry \citep[e.g.][]{mozurkewich03,boyajian12,boyajian12b,boyajian13}. Figure \ref{fig:huang} again shows good agreement, with a residual median and scatter of 109\,K and 173\,K, respectively. Both comparisons show that our temperatures are on average $\sim$\,100\,K hotter, which is comparable to previously found offsets between temperature scales \citep{pinsonneault11} and well within the systematic uncertainty of the fundamental interferometric temperature scale itself \citep[e.g.][]{white18}. Based on these comparisons we have adopted a conservative uncertainty of 3\% on the temperatures in the ATL, which encompasses both random and systematic uncertainties from the literature comparisons.

%%%%%%%%%%%%%%%%%%%%%%%%%%%%%%%%%%%%%%%%%%%%%%%%%%%%%%%%%%%%%%%%%%%%%%%%%%%%%%%%%%%%%%%%%%%%%%%%%%%%%

%%%%%%%%%%%%%%%%%%%%%%%%%%%%%%%%%%%%%%%%%%%%%%%%%%%%%%%%%%%%%%%%%%%%%%%%%%%%%%%%%%%%%%%%%%%%%%%%%%%%%

%%%%%%%%%%%%%%%%%%%%%%%%%%%%%%%%%%%%%%%%%%%%%%%%%%%%%%%%%%%%%%%%%%%%%%%%%%%%%%%%%%%%%%%%%%%%%%%%%%%%%

%%%%%%%%%%%%%%%%%%%%%%%%%%%%%%%%%%%%%%%%%%%%%%%%%%%%%%%%%%%%%%%%%%%%%%%%%%%%%%%%%%%%%%%%%%%%%%%%%%%%%

\begin{figure}
	\centering
    \resizebox{\hsize}{!}{\includegraphics{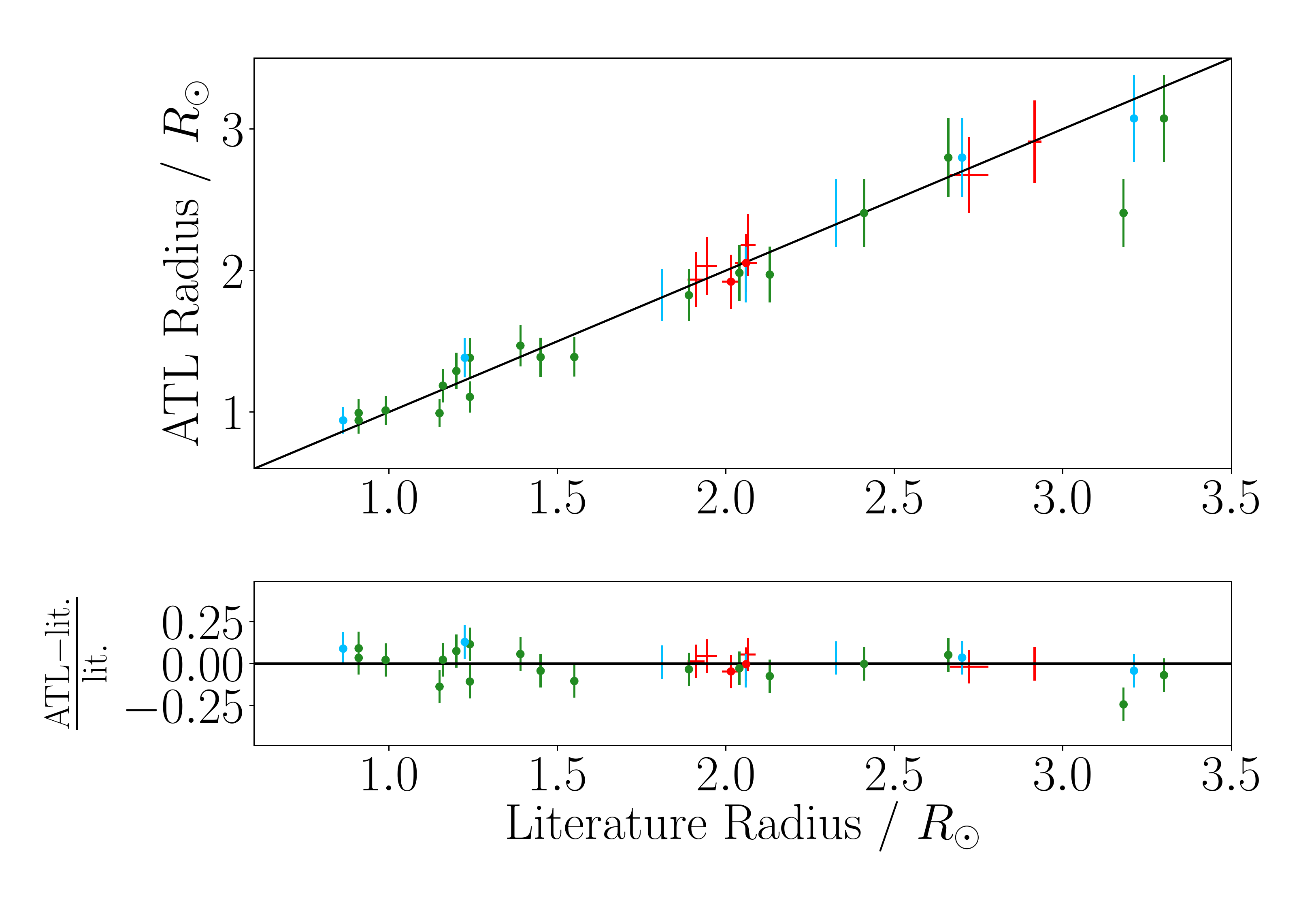}}
	\caption{Comparison between the literature radii derived from parallaxes (red and blue symbols) and interferometry (green symbols). We observe good agreement, with a residual median and scatter of 0.04\,\% and 0.07\,\%.}	
	\label{fig:ag_radii}
\end{figure}

Next, we compared radii in the ATL to a selection of bright stars in \citet{silva_aguirre_verifying_2012} and \citet{bruntt_accurate_2010}. \citet{silva_aguirre_verifying_2012} derived radii for a small number of \emph{Kepler} solar-type stars that have detections of solar-like oscillations as well as precise \emph{Hipparcos} parallaxes. \citet{bruntt_accurate_2010} estimated the radii of even brighter stars using two approaches: first, using measurements of limb-darkened stellar angular diameters and stellar parallaxes; and second, using the Stefan-Boltzmann law with luminosities derived from $V$-band magnitudes, bolometric corrections and parallaxes, and spectroscopic temperatures, i.e., the basic approach we have used but with some different observables. Figure~\ref{fig:ag_radii} shows the comparison with between ATL and those literature values. We observe excellent agreement, with a residual median and scatter of 0.04\,\% and 0.07\,\%, respectively. Overall, these comparisons confirm that the stellar properties derived in the ATL do not suffer from large systematic errors when compared with literature values.

%%%%%%%%%%%%%%%%%%%%%%%%%%%%%%%%%%%%%%%%%%%%%%%%%%%%%%%%%%%%%%%%%%%%%%%%%%%%%%%%%%%%%%%%%%%%%%%%%%%%%%%%%%

%%%%%%%%%%%%%%%%%%%%%%%%%%%%%%%%%%%%%%%%%%%%%%%%%%%%%%%%%%%%%%%%%%%%%%%%%%%%%%%%%%%%%%%%%%%%%%%%%%%%%%%%%%

%%%%%%%%%%%%%%%%%%%%%%%%%%%%%%%%%%%%%%%%%%%%%%%%%%%%%%%%%%%%%%%%%%%%%%%%%%%%%%%%%%%%%%%%%%%%%%%%%%%%%%%%

%\begin{figure}
%	\centering
%	\includegraphics[scale=0.8]{bruntt_radii_ld.pdf}
%	\caption{Comparison between the \citet{bruntt_accurate_2010} radii from the 'limb-darkening' method (using the Steffan-Boltzmann law) and the ATL. The standard deviation is given in the top left corner of the plot. The two sets of radii agree well.}	
%	\label{fig:bruntt_ld}
%\end{figure}

%\begin{figure}
%	\centering
%	\includegraphics[scale=0.8]{bruntt_radii_dir.pdf}
%	\caption{Comparison between \citet{bruntt_accurate_2010} radii from the 'direct' method (using the Steffan-Boltzmann law) and the ATL. The standard deviation is given in the top left corner of the plot. The two sets of radii agree well.}	
%	\label{fig:bruntt_dir}
%\end{figure}

%%%%%%%%%%%%%%%%%%%%%%%%%%%%%%%%%%%%%%%%%%%%%%%%%%%%%%%%%%%%%%%%%%%%%%%%%%%%%%%%%%%%%%%%%%%%%%%%%%%%%%%%

\section{ATL Construction}
\label{sec:construction}

\subsection{Consolidation of DR2 and XHIP entries}

\begin{figure}
	\centering
    \resizebox{\hsize}{!}{\includegraphics{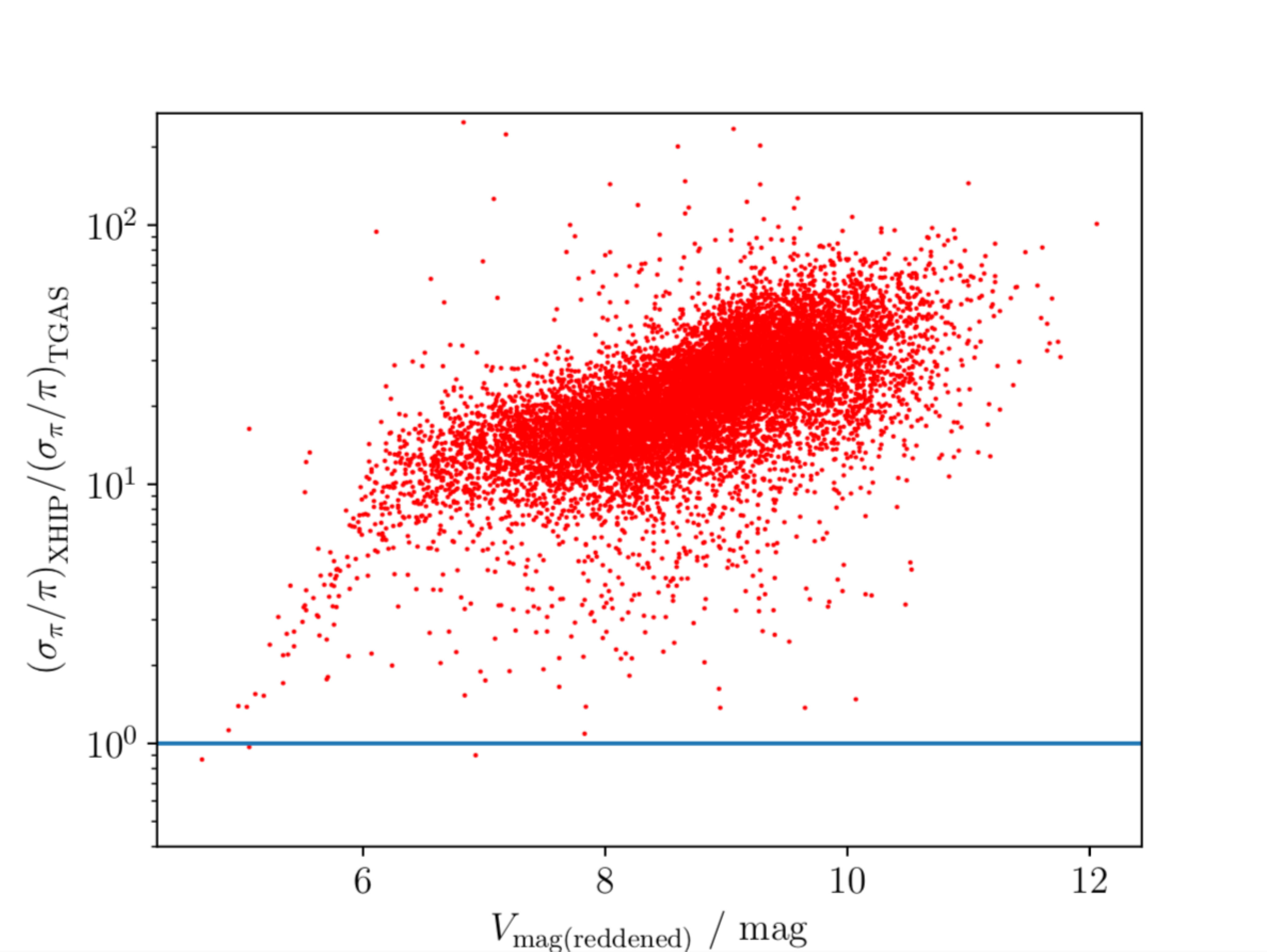}}
	\caption{The stars with both DR2 and XHIP entries. The fractional parallax $\sigma_{\pi}/\pi$ value was calculated from the DR2 and XHIP entries for each star separately. The parameters from the catalogue with the lower $\sigma_{\pi}/\pi$ value was chosen. XHIP properties were only used for the few stars below the blue line.}	
	\label{fig:frac plx}
\end{figure}

Having removed stars with large fractional parallax uncertainties (see Section 3.2), we combined the retained stars from DR2 and XHIP into a single list to be treated homogeneously. This combined list contained over 300,000 stars. Most had entries in the DR2 catalog, with only a small number in XHIP. However, there were $\sim$\,17,000 stars which existed in both lists. We broke this degeneracy by using data and derived parameters from the catalog whose target entry had the smaller fractional parallax uncertainty of the two. Not surprisingly, in the vast majority of cases the DR2 entries were selected, with only a handful of bright XHIP targets being retained where \textit{Hipparcos} outperforms \textit{Gaia} (see Figure \ref{fig:frac plx}).

%%%%%%%%%%%%%%%%%%%%%%%%%%%%%%%%%%%%%%%%%%%%%%%%%%%%%%%%%%%%%%%%%%%%%%%%%%%%%%%%%%%%%%%%%%%%%%%%%%%%%%%%%%%

%%%%%%%%%%%%%%%%%%%%%%%%%%%%%%%%%%%%%%%%%%%%%%%%%%%%%%%%%%%%%%%%%%%%%%%%%%%%%%%%%%%%%%%%%%%%%%%%%%%%%%%%%%

\subsection{Down-selection to solar-like oscillating short-cadence targets}

From the above combined list, we selected targets that are potential solar-like oscillators. To do this, we retained all stars that lie on the cool side (redwards) of the $\delta$\,Scuti instability strip, i.e., those having $T_{\rm eff} < T_{\rm red}$, with the red-edge temperature defined as \citet{chaplin_predicting_2011}:
\begin{equation}
\label{eq:instab}
T_{\rm red} = 8907{\rm \,K} \times (L/L_{\odot})^{-0.093}.
\end{equation}

We further restricted to targets that have predicted dominant oscillation frequencies requiring the TESS short-cadence (2-min) data. Solar-like oscillators present a rich spectrum of detectable overtones, with oscillation power following a Gaussian-like envelope centered on the so-called frequency of maximum oscillations power, $\nu_{\rm max}$. We retained all targets having $\nu_{\rm max} \geqslant 240\,\rm \mu Hz$. This represents, to reasonable approximation, an upper-limit cut in luminosity that discards low-luminosity red-giants at or just above the base of the red-giant branch, i.e., giants whose solar-like oscillations can be very readily resolved in the TESS 30-min long-cadence FFI data. The $240\,\rm \mu Hz$ limit was set deliberately to lie below the FFI Nyquist frequency of $278\,\rm \mu Hz$ to account for uncertainties in the ATL-based predictions and also to provide a reasonable sample of targets in short-cadence whose oscillation spectra are reasonably close to the Nyquist limit. Experience from \emph{Kepler} has shown that such spectra can be difficult to analyze using long-cadence data only, due to aliasing about the Nyquist frequency \citep[e.g.][]{yu16}.

\begin{figure}
	\centering
    \resizebox{\hsize}{!}{\includegraphics{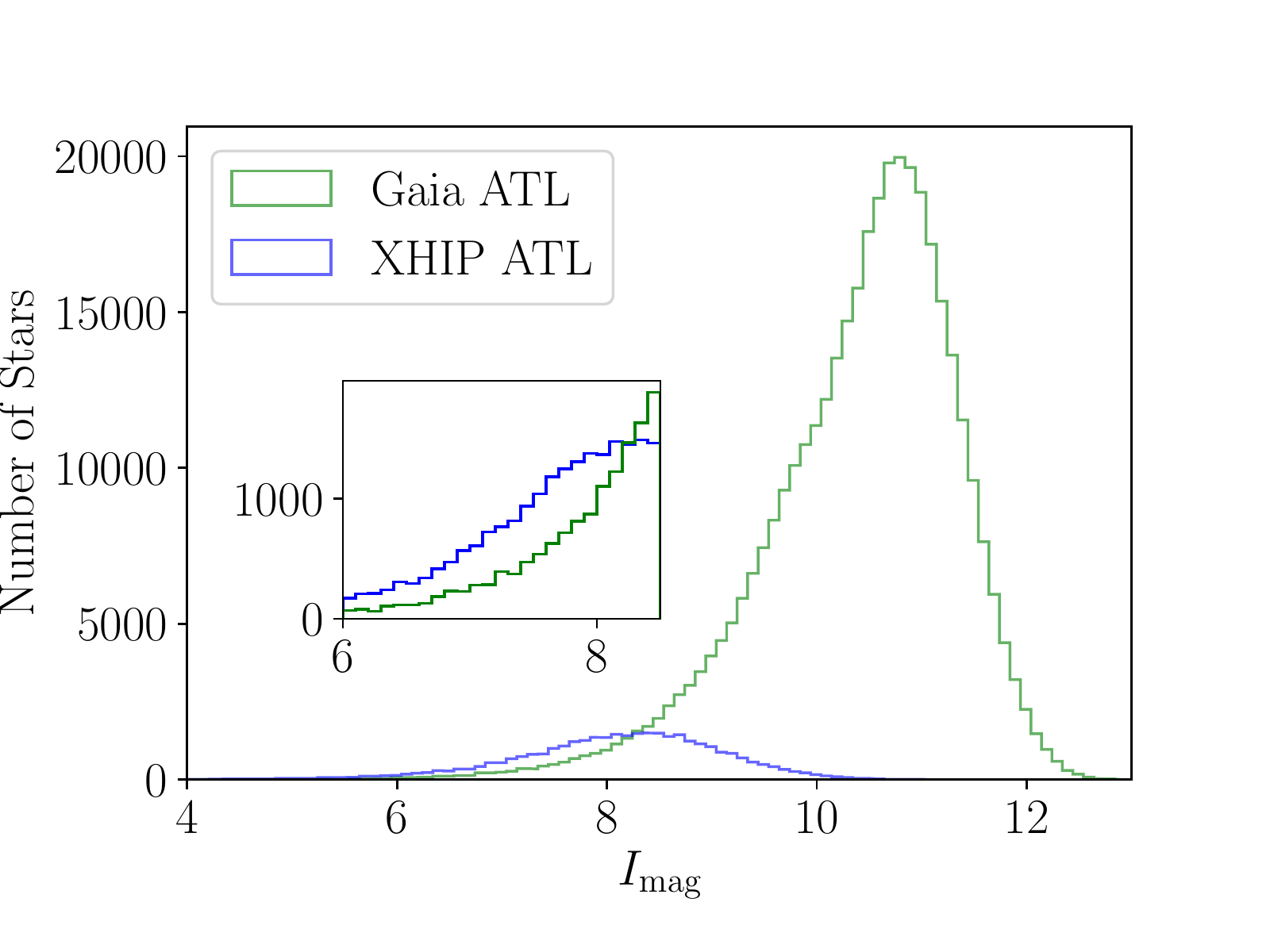}}
	\caption{The $I_{\rm mag}$ distribution of the XHIP and DR2 catalogues. Shown here after \teff/$L$ cuts, but before $P_{\rm det}$ was calculated. The inset shows the \imag region where the two catalogues overlap.}	
	\label{fig:imag hist}
\end{figure}

The boundary in the $L$-$T_{\rm eff}$ plane for the $\nu_{\rm max}$ cut follows from the approximate relation (see \citealt{campante_asteroseismic_2016}):
\begin{equation}
\label{eq:numax}
\nu_{\rm max} = \nu_{\rm max,\odot} \left( \frac{R}{R_{\odot}} \right)^{-1.85}
                                    \left( \frac{T_{\rm eff}}{T_{\rm eff,\odot}} \right)^{0.92},
\end{equation}
which, combined with $L \propto R^2T_{\rm eff}^4$ and setting $\nu_{\rm max} = 240\,\rm \mu Hz$, defines the boundary
\begin{equation}
L/\lsol \leqslant 16.7 \times \left( \frac{T_{\rm eff}}{T_{\rm eff,\odot}} \right)^5
\end{equation}
for retaining targets. Figure \ref{fig:imag hist} shows the $I$ magnitude distribution of the Hipparcos and Gaia subsamples of the ATL after these H-R diagram cuts have been applied. As expected, Gaia dominates the faint end of the ATL, and the drop-off at $I \sim 11$ is caused by the fractional parallax precision cut described in Section 3.2.

%%%%%%%%%%%%%%%%%%%%%%%%%%%%%%%%%%%%%%%%%%%%%%%%%%%%%%%%%%%%%%%%%%%%%%%%%%%%%%%%%%%%%%%%%%%%%%%%%%%%%%%%%%

%%%%%%%%%%%%%%%%%%%%%%%%%%%%%%%%%%%%%%%%%%%%%%%%%%%%%%%%%%%%%%%%%%%%%%%%%%%%%%%%%%%%%%%%%%%%%%%%%%%%%%%

\subsection{Estimation of asteroseismic detection probabilities}

To calculate asteroseismic detection probabilities we used the approach developed by \citet{chaplin_predicting_2011}, which has been applied successfully to short-cadence target selection for \emph{Kepler} Objects of Interest in the \emph{Kepler} nominal mission and, more recently, to short-cadence target selection for solar-type stars observed with K2 (e.g. see \citealt{Chaplin2015, Lund2016a, Lund2016b}). The approach is based on predicting the global signal-to-noise ratio in the oscillation spectrum, i.e., the predicted total power in the observed solar-like oscillations divided by the total power from granulation, shot and instrumental noise, summed across the range in frequency occupied by the modes. The total oscillation and granulation power across the frequency range of interest centered on the predicted $\nu_{\rm max}$ may be calculated from the previously derived $L$, $T_{\rm eff}$ and $R$. The shot and instrumental noise depend on the instrumental performance and the apparent magnitude of targets in the instrumental bandpass. The duration of the observations is also an important factor: at a given global signal-to-noise ratio, the detection probability will rise as the length of observations is increased. 

The formulation by \citet{chaplin_predicting_2011} was updated for the TESS instrumental specifications in \citet{campante_asteroseismic_2016}. We followed that revised recipe in detail here, and refer the reader to Section~3 of \citet{campante_asteroseismic_2016} for the relevant steps and relations. We have made some changes to the estimation of the TESS noise, to reflect updates to information that is available on the instrumental performance. We describe those small changes next in Section~\ref{sec:noise}. 

%We then go on to describe in detail in Section~\ref{sec:tobs} how the expected duration of TESS observations was calculated for every potential target. This was not covered in \citet{campante_asteroseismic_2016} and there are some subtleties relevant to construction of the ATL that need to be discussed in detail here.

\subsubsection{Updates to noise predictions}
\label{sec:noise}

The predicted instrumental noise is dominated by the shot noise, but also includes contributions to represent contamination from nearby stars, and readout noise. Since the ATL targets are bright, contamination is expected to be modest, in spite of the large point-spread function of TESS. Note that we assumed that the systematic noise floor of 60\,ppm per hour \citep{sullivan_transiting_2015} is negligible, since this is a design threshold requirement for meeting core exoplanet science deliverables and will not reflect the actual performance.

\begin{figure}
	\centering
    \resizebox{\hsize}{!}{\includegraphics{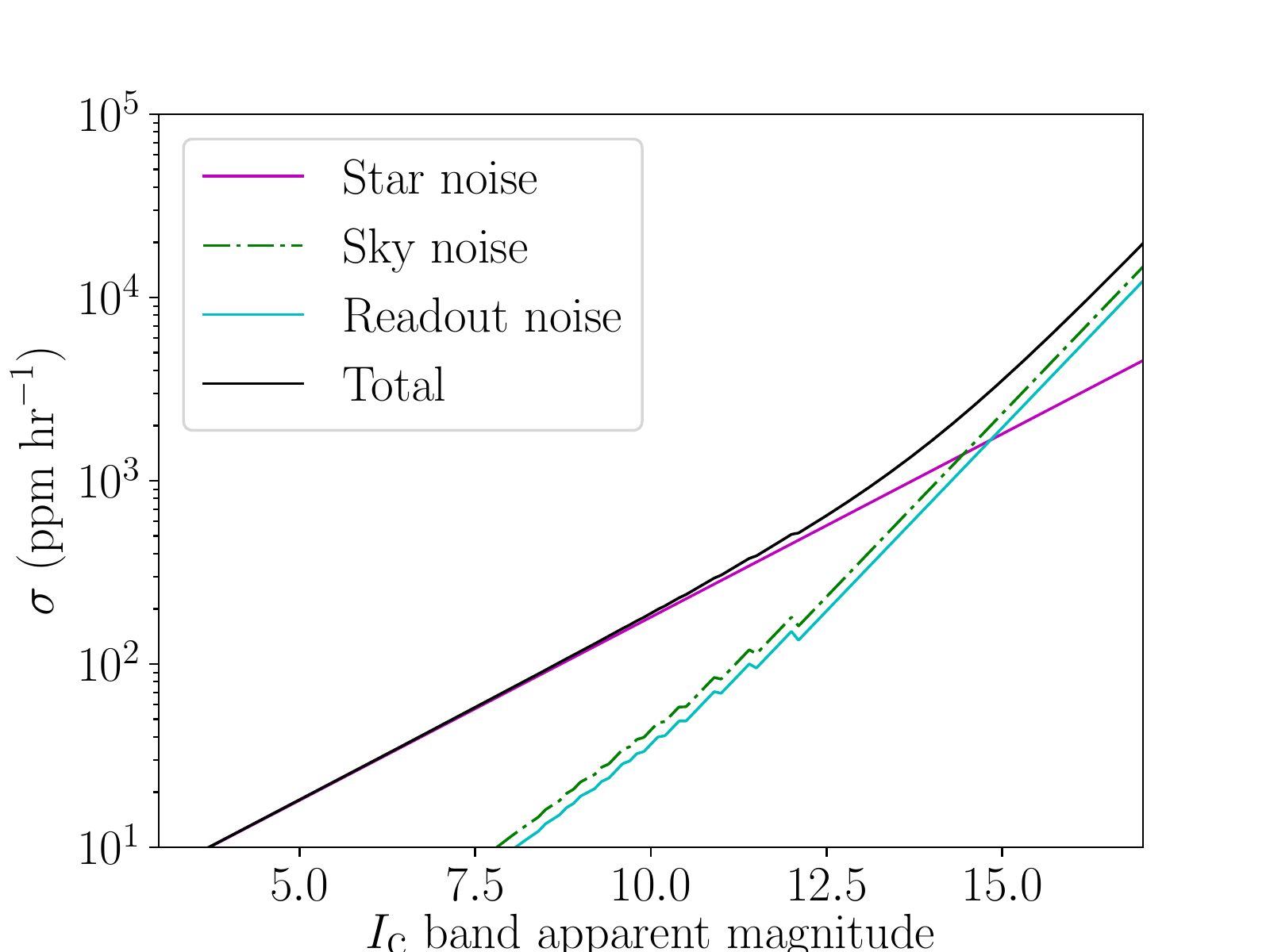}}
	\caption{Individual noise contributions (coloured lines) and total noise budget (black line) as a function of apparent $I$ magnitude used to calculate asteroseismic detection probabilities.}
	\label{fig:noise}
\end{figure}

%\textbf{Tiago/Mat: What follows may not be entirely correct... please check and update as necessary... Mat: I've checked, it looks good to me.}
As in \citet{campante_asteroseismic_2016}, we used the \texttt{calc\_noise} IDL procedure provided by the TESS Science Team \citep{sullivan_transiting_2015} to calculate the instrumental noise, which takes the $I$-band magnitude as its main input. There are two updates: (i) the absolute calibration of the expected noise levels is now slightly higher, due to a reduced estimated effective aperture size for the instrument; and (ii) the expected number of pixels, $N_{\rm mask}$, in each stellar pixel mask is now smaller, which has the effect of reducing noise levels. Updated mask sizes were calculated using the simple parametric model provided by the TESS team (J. Winn, private communication):
 \begin{equation}
 \label{eq:pixel cost}
 N_{\rm mask} =  10^{0.8464 - 0.2144 \times (I_{\rm mag} - 10.0)},
 \end{equation}
and the number of pixels was rounded up to the nearest whole number. Once calculated, the individual instrumental noise contributions (see Figure \ref{fig:noise}) were summed in quadrature to give the total instrumental noise per 2-min cadence.

%%%%%%%%%%%%%%%%%%%%%%%%%%%%%%%%%%%%%%%%%%%%%%%%%%%%%%%%%%%%%%%%%%%%%%%%%%%%%%%%%%%%%%%%%%%%%%%%%%%%%%%%

%%%%%%%%%%%%%%%%%%%%%%%%%%%%%%%%%%%%%%%%%%%%%%%%%%%%%%%%%%%%%%%%%%%%%%%%%%%%%%%%%%%%%%%%%%%%%%%%%%%%%%%

\subsubsection{The Observation time in the TESS field of view}
\label{sec:tobs}

We begin this section with a recap of basic information on the TESS field of view and observing strategy (see also, e.g., \citealt{ricker_transiting_2014, sullivan_transiting_2015, huang18}). TESS comprises four CCD cameras. Each CCD images a $24^{\circ}\times24^{\circ}$ area on the sky, with the total collecting area of the four cameras at any given time being a strip of dimensions $24^{\circ}$ (ecliptic longitude) $\times$ $96^{\circ}$ (ecliptic latitude). TESS will survey the sky south of the ecliptic in its first year of science operations, with the hemisphere divided into 13 strips. Each resulting \emph{sector} pointing will last, on average, about 27.4\,days. The durations of each sector pointing differ by up to 1.5\,days due to variations in the length of the spacecraft's orbit. The sky north of the ecliptic will be observed in the second year of nominal science operations.

\iffalse
\begin{figure}[htbp]
	\centering
	\includegraphics[scale=0.3]{regions}
	\caption{The 4 CCD cameras image a total area on the sky of $24^{\circ}$ (longitude) $\times$ $96^{\circ}$ (latitude). The CCD camera that will image the part of the sky at the highest latitude is centred around the ecliptic pole. TESS will observe 13 regions in one hemisphere before observing 13 regions in the other hemisphere. Image from the NASA Transiting Exoplanet Survey Satellite website.}	
	\label{fig:regions}	
\end{figure}
\fi

\begin{figure}[htbp]
	%\begin{minipage}[t]{0.45\textwidth}
	\centering
    %\resizebox{7.5cm}{!}{\includegraphics{TESSfield09_07}}
    %\resizebox{8cm}{!}{\includegraphics{task_5}}
	%\includegraphics[height=60mm]{TESSfield09_07}
    %\includegraphics[height=60mm]{task_5}
	%\caption{The TESS field of view on the celestial sphere. The black lines are lines of constant longitude or latitude. The colourbar represents how long a part of the sky will be observed by TESS. Different regions of observation will overlap with each other so stars in certain parts of the sky will be imaged for longer. Stars close to the northern and southern ecliptic poles will be imaged for up to 351 days (they will lie in 13 sectors of observation). }
	\label{fig:field1}
	%\end{minipage}\hfill
	%\begin{minipage}[t]{0.45\textwidth}
	%\centering
	%\includegraphics[height=60mm]{task_5}
    \resizebox{\hsize}{!}{\includegraphics{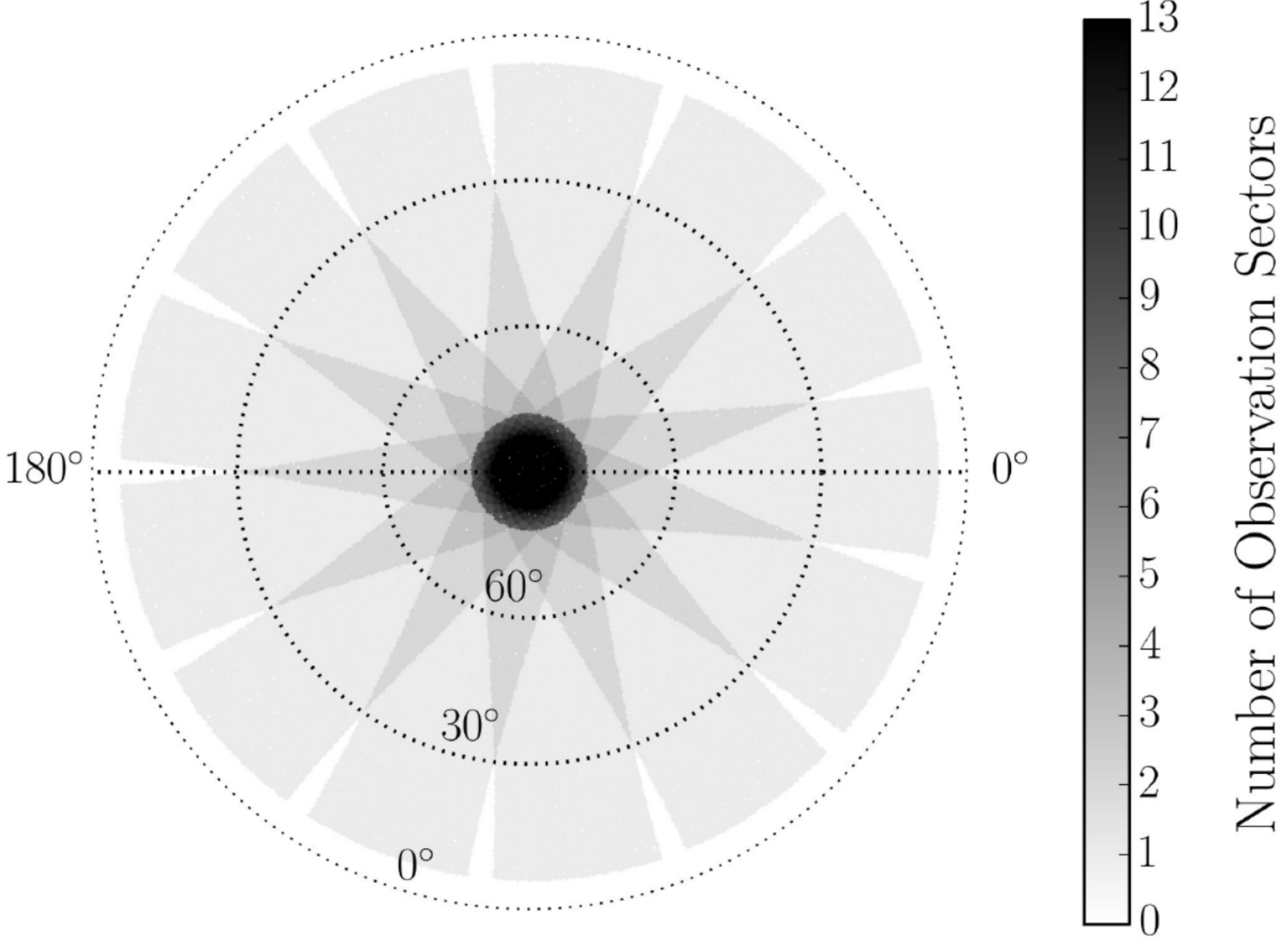}}
	\caption{The field of view from one of the ecliptic poles. The colourbar represents how long a part of the sky will be observed by TESS. The dotted circles are lines of constant latitude ($0^{\circ}$,  $30^{\circ}$ and $60^{\circ}$). The centre of the image has a latitude of $90^{\circ}$. The outer dotted circle at $0^{\circ}$ latitude has labels for $0^{\circ}$ and $180^{\circ}$ longitude. The horizontal dotted line represents longitude values of $0^{\circ}$ and $180^{\circ}$.}
	\label{fig:field2}	
	%\end{minipage}
\end{figure}

The majority of TESS targets will be observed over only one 27-day sector. The duration increases for latitudes significantly above or below the ecliptic plane, because targets may then be observed in more than one sector pointing, reaching a maximum of 13 sectors, i.e., about 351 days, at the ecliptic poles (the Continuous Viewing Zone, see Figure \ref{fig:field2}). TESS will not observe targets within $\pm 6^{\circ}$ of the ecliptic during its nominal mission, and those stars were removed from our list.

Figure \ref{fig:field2} shows that there will be small gaps between sectors. At the time the ATL was delivered, the initial pointing at the commencement of science operations  was not known.  As can be seen from the figures, that will influence not only which stars are
missed by TESS (i.e., those falling in the sector-to-sector gaps at low ecliptic latitudes) but also
the numbers of sectors for which targets at higher latitudes will be observed. Here, we ignore the
low-latitude gaps and assume that all stars with ecliptic latitudes beyond $\pm 6^{\circ}$ are potentially observable. Targets that fall in the gaps will be discarded when the TESS team compiles actual target lists for each known pointing.

For higher-latitude targets, there are several options open to us. We could adopt a particular pointing and then compute the resulting number of observation sectors for each target for input to the asteroseismic detection recipe. We could instead estimate the minimum and maximum potential number of observation sectors for each target, which depend on the ecliptic latitude but not the exact pointing, and use one or the other as input to the detection recipe. While this choice will affect the rank ordering of targets based on the detection probability, it turns out that the resulting changes in ranking are typically a few hundred places or less, a change that is very unlikely to influence whether targets with potentially detectable oscillations are observed by TESS. As such we adopted the simpler first option and assumed an initial pointing of $E_{\rm long} = 0^{\circ}$. 

%%%%%%%%%%%%%%%%%%%%%%%%%%%%%%%%%%%%%%%%%%%%%%%%%%%%%%%%%%%%%%%%%%%%%%%%%%%%%%%%%%%%%%%%%%%%%%%%%%%%%%%%%

%%%%%%%%%%%%%%%%%%%%%%%%%%%%%%%%%%%%%%%%%%%%%%%%%%%%%%%%%%%%%%%%%%%%%%%%%%%%%%%%%%%%%%%%%%%%%%%%%%%%%%

\begin{figure}[htbp]
	\centering
	\includegraphics[scale=1.3]{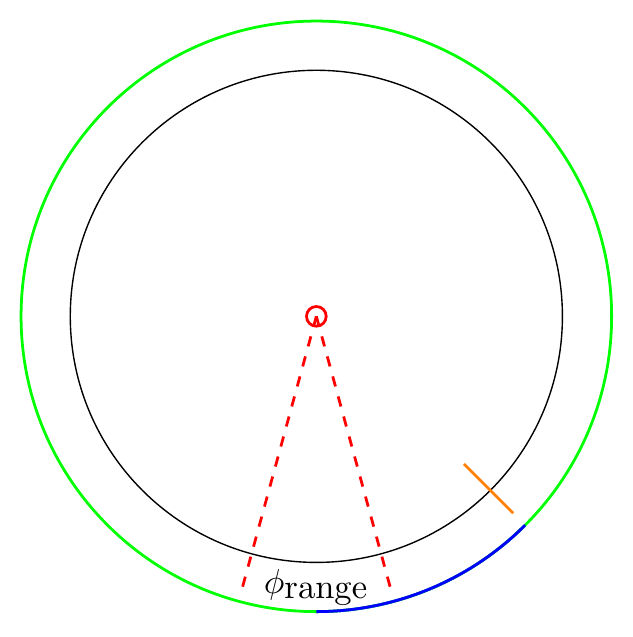}
	\caption{An example of what is calculated in order to determine whether a star lies inside $\phi_{\textrm{range}}$. In this example, the star's longitude (represented by the orange line) lies outside of the satellite's field of view marked by the red dashed lines (Equation \ref{eq:range}).}
	\label{fig:fov}	
\end{figure}

The ecliptic position ($E_{\rm long}, E_{\rm lat}$) determines how long a star can be observed. 
To determine whether a star is observable in any given sector pointing we must define the longitudes of the center of each observing sector, $E_{\rm CCD}$, and the longitude range, $\phi_{\rm range}$, that the cameras cover at a given latitude (i.e., the latitude of the star). Figure \ref{fig:fov} gives a pictorial representation of $\phi_{\rm range}$.  The black circle in Figure \ref{fig:fov} is a line of constant latitude. In Figure \ref{fig:fov}, the satellite is represented by the small red circle in the center of the image. The red dashed lines show the width of the field-of-view of TESS. These are the edges of $\phi_{\rm range}$. 

$\phi_{\rm range}$ is given by
\begin{equation}
\label{eq:range}
\phi_{\rm range}  = \frac{24^{\circ}}{\textrm{cos}(E_{\rm lat})},
\end{equation}
where $E_{\rm lat}$ is the latitude of the star in question and $24^{\circ}$ is the width of the field covered by the CCD cameras at $0^{\circ}$ latitude. If the longitude of the star lies within $\pm\phi_{\rm range} / 2$ then the image of the star will be captured by a camera. In order to check this, the difference between the center of the CCD ($E_{\rm CCD}$) and the longitude of the star ($E_{\rm long}$) must be calculated. This difference is given by
\begin{equation}
 \phi_{\rm Diff} = \left\{ \,
    \begin{IEEEeqnarraybox}[][c]{l?s}
      \IEEEstrut
      |E_{\rm CCD} - E_{\rm long}| , \\
      360^{\circ} - |E_{\rm CCD} - E_{\rm long}| .
      \IEEEstrut
    \end{IEEEeqnarraybox}
\right.
  \label{eq:long_diff}
\end{equation}

Equation \ref{eq:long_diff} will produce two values of $\phi_{\rm Diff}$, as shown by the blue and green lines in Figure \ref{fig:fov}. Only the smaller distance between $E_{\rm long}$ and $E_{\rm CCD}$ should be taken as the distance between the star and the center of the field-of-view. The longitudinal position of a star is marked in Figure \ref{fig:fov} by the orange line. %The green and blue arcs show the two values of $\phi_{\rm Diff}$ calculated from equation \ref{eq:long_diff}.

Now, if $\phi_{\rm range} \geqslant \phi_{\rm Diff}$, the star will be observed in that region. In Figure \ref{fig:fov}, the length of the blue arc will be the accepted value for $\phi_{\rm Diff}$, since it is shorter than the green arc. However, although the blue arc is the shorter of the two, $\phi_{\rm range} < \phi_{\rm Diff}$ and so the star will not be observed in this sector.

%%%%%%%%%%%%%%%%%%%%%%%%%%%%%%%%%%%%%%%%%%%%%%%%%%%%%%%%%%%%%%%%%%%%%%%%%%%%%%%%%%%%%%%%%%%%%%%%%%%%%

%%%%%%%%%%%%%%%%%%%%%%%%%%%%%%%%%%%%%%%%%%%%%%%%%%%%%%%%%%%%%%%%%%%%%%%%%%%%%%%%%%%%%%%%%%%%%%%%%%%%%%

Using Equations \ref{eq:range} and \ref{eq:long_diff}, we determined which stars would be observable in the first sector pointing of each ecliptic hemisphere, again taking each to be centered on $E_{\rm CCD}=0^{\circ}$. The same calculations were then repeated for each subsequent pointing, with every adjacent pointing shifted by  $E_{\rm CCD}=360^{\circ}/13=27.7^{\circ}$. The observing time $T_{\rm obs}$ was then obtained from the maximum contiguous number of sectors that each star is observed.

\subsubsection{Rank-ordering the ATL using the detection probabilities}
\label{sec:rank}

% DH: rewrite of this section:

\begin{figure*}
	\includegraphics[scale=0.55]{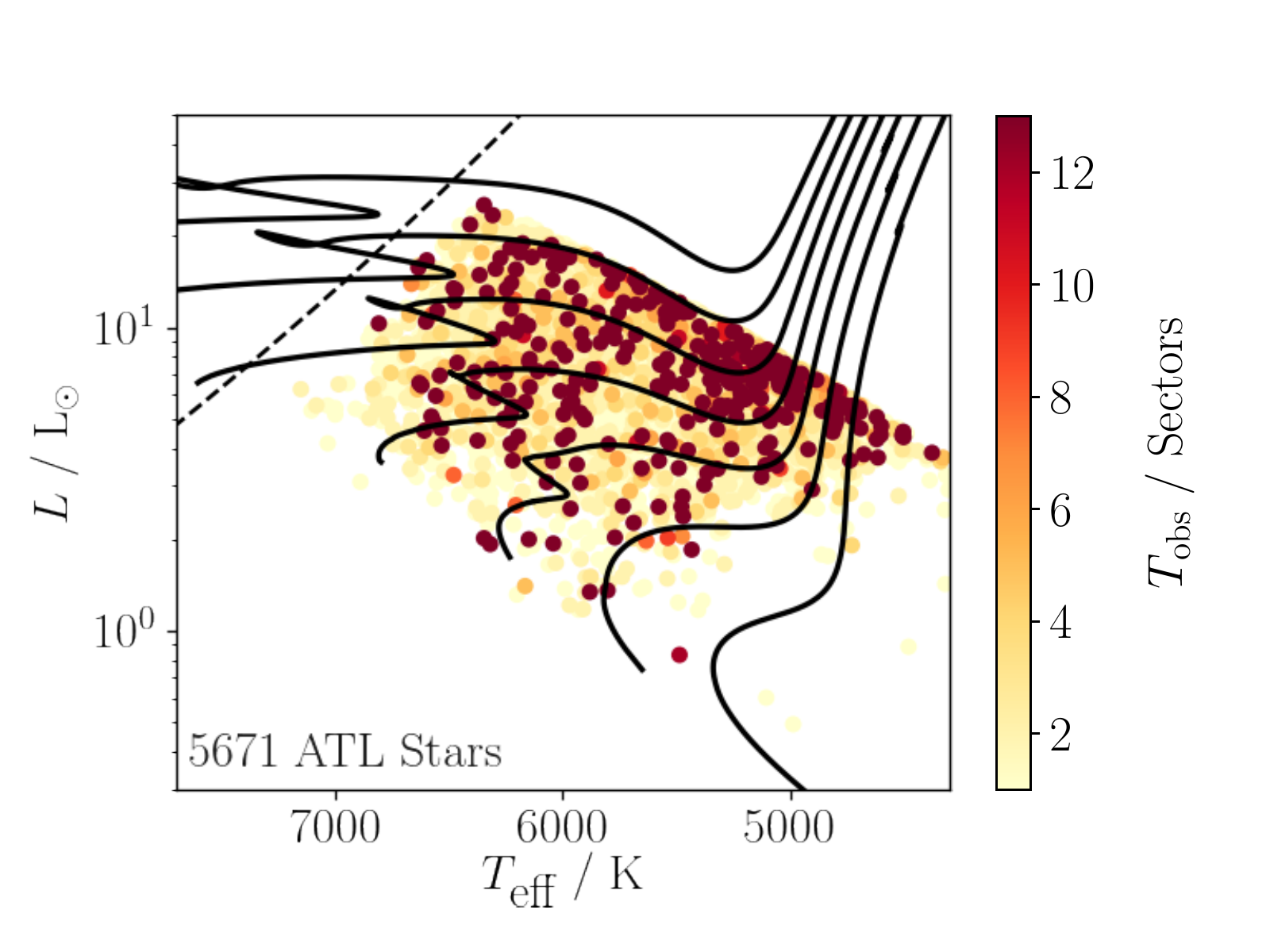}
	\includegraphics[scale=0.55]{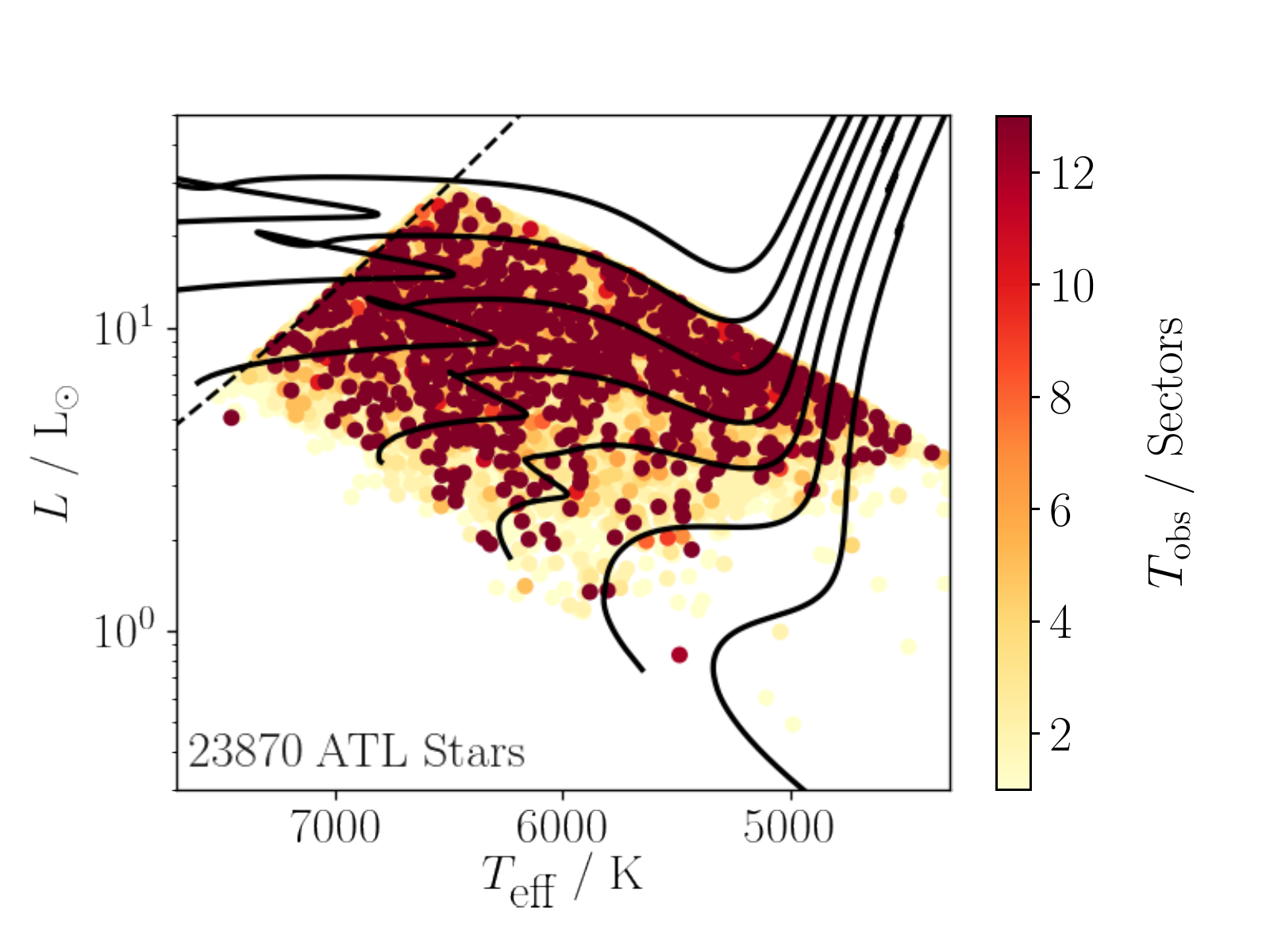}
    
	\caption{H-R diagram of all ATL stars with detection probabilities greater than 50\,\% with (left panel) and without (right panel) including the $\beta$ factor, which accounts for the attenuation of oscillation amplitudes towards the red edge of the instability strip (dashed line). Solid lines show solar metallicity evolutionary tracks with masses from $0.8\,M_{\odot}$ to $2.0\,M_{\odot}$ in steps of $0.2\,M_{\odot}$. Note that the sharp edges are due to cuts at the red edge of the instability strip (Equation~\ref{eq:instab}) and stars oscillating with frequencies accessible with TESS FFI data (Equation~\ref{eq:numax}).}
	\label{fig:pdets}
\end{figure*}

The analysis of the \emph{Kepler} sample demonstrated that amplitudes of solar-like oscillations are progressively reduced relative to predictions from scaling relations when moving from late to early F-type stars. \citet{chaplin_predicting_2011} attempted to explicitly capture this effect by introducing an attenuation factor $\beta$, which was also adopted by \citet{campante_asteroseismic_2016} to describe the maximum amplitude for radial mode oscillations in the TESS bandpass:
 \begin{equation}
 \label{eq:amp}
 A_{\rm max} = 0.85\times2.5{\rm \,ppm} \times \beta \times \left(\frac{R}  {R_{\odot}}\right)^{1.85}\times \left(\frac{T_{\rm eff}}{T_{\rm eff,\odot}}\right)^{0.57}.
 \end{equation}
Here, $\beta$ is given by:
 \begin{equation}
 \label{eq:beta}
 \beta = 1.0 - \exp \left[ (T_{\rm red} - T_{\rm eff})/1550\,\mathrm{K} \right],
 \end{equation}
where $T_{\rm red}$ is the previously defined temperature on the red-edge of the $\delta$\,Scuti instability strip at the luminosity of the target (see Equation~\ref{eq:instab}). The attenuation given by $\beta$ reduces predicted mode amplitudes in hotter stars, and hence lowers detection probabilities and the associated rank-ordering of those targets. Figure \ref{fig:pdets} illustrates the effect by showing all ATL stars with detection probabilities greater than 50\,\% with and without including the $\beta$ factor. As expected, the $\beta$ factor strongly reduces the number of stars with significant detection probabilities, especially towards the instability strip.

Rank-ordering the ATL using the detection probabilities including the $\beta$ factor would optimize the yield of asteroseismic detections with TESS. However, using the $\beta$ factor would also strongly bias against making new discoveries in stars that do not fit the trend in asteroseismic amplitudes shown by the \emph{Kepler} sample on which the detection recipe is based (which is, by definition, an already biased sample).  

The group most affected by this comprises hot F-type stars, which lie at the boundary where solar-like oscillations diminish to undetectable amplitudes and classical pulsations driven by the $\kappa$ mechanism start to become excited. Determining the details of this transition is of considerable interest for understanding the driving and damping of oscillations, and intriguing examples of ``hybrid stars'' showing signatures of solar-type oscillations and classical pulsators have already been detected \citep{kallinger10,antoci11}, leading to suggestions of new pulsation driving mechanisms \citep{antoci14}. The sampling of targets in this region was sparse for \emph{Kepler}, and limited by the small number of short-cadence target slots available at any one time. There is now the potential to address those issues with TESS.

To mitigate the strong bias against hot stars in the ATL we define a new probability, $p_{\rm mix}$, as follows:
 \begin{equation}
 p_{\rm mix} = (1-\alpha)\,p_{\rm vary} + \alpha\,p_{\rm fix}.
 \label{eq:pmix}
 \end{equation}
Here, $p_{\rm vary}$ is the detection probability calculated using the $\beta$ factor, $p_{\rm fix}$ is the detection probability calculated by fixing $\beta = 1$ for all stars (i.e. ignoring amplitude attenuation), and $\alpha$ regulates the relative weighting between $p_{\rm vary}$ and $p_{\rm fix}$. After investigating the rank ordered lists using a range of values of $\alpha$, we found that $\alpha=0.5$ (i.e. equal weighting between $p_{\rm vary}$ and $p_{\rm fix}$) provides the best overall compromise between obtaining a significant yield and including enough hot stars at high ranks. For the remainder of the paper, all ranked lists in the ATL were calculated with detection probabilities using $\alpha = 0.5$.

\begin{figure*}
	\centering
	\includegraphics[width=18.0cm]{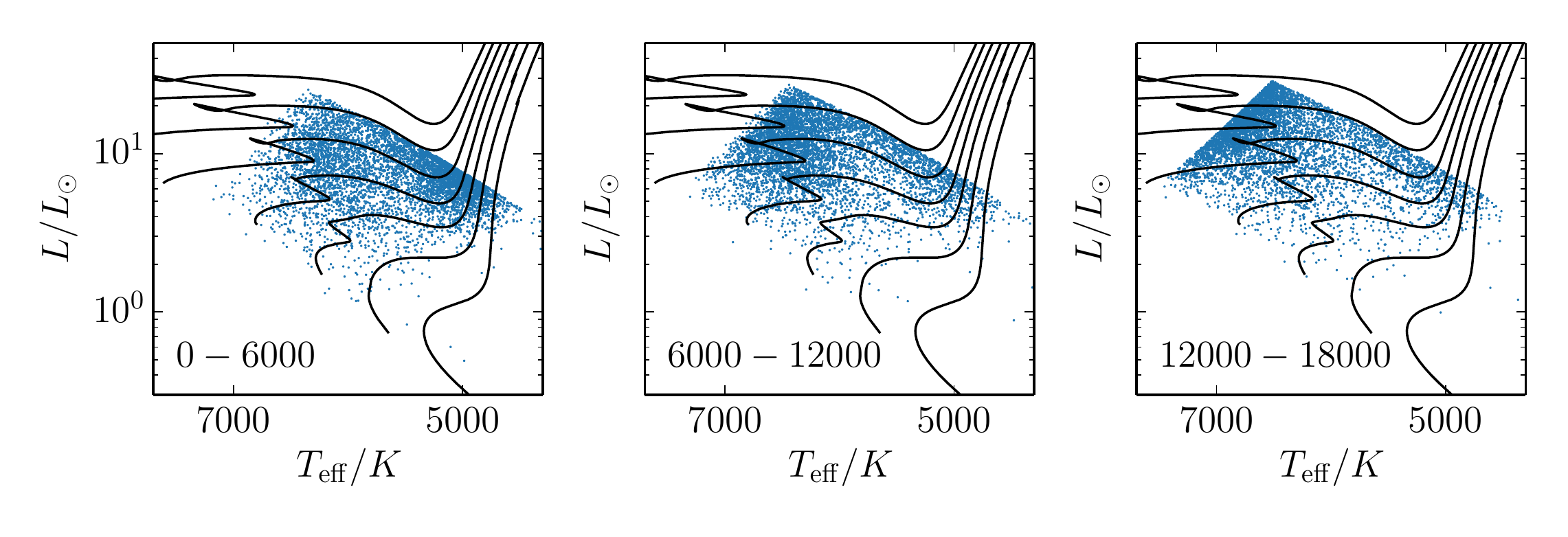}
	\caption{H-R diagram of stars in the ATL. Each panel shows six thousand stars according to their ranking. Black lines show evolutionary tracks with masses from $0.8\,M_{\odot}$ to $2\,M_{\odot}$ in steps of $0.2\,M_{\odot}$.}	
	\label{fig:hrd}
\end{figure*}

%%%%%%%%%%%%%%%%%%%%%%%%%%%%%%%%%%%%%%%%%%%%%%%%%%%%%%%%%%%%%%%%%%%%%%%%%%%%%%%%%%%%%%%%%%

\section{Overview of the Asteroseismic Target List}
\label{sec:overview}

\subsection{Distribution across H-R Diagram}

Figure \ref{fig:hrd} shows the distribution of the 18,000 top-ranked stars in the ATL in an H-R diagram, split into bins of 6000 stars each. Similar to Figure \ref{fig:pdets}, the sharp edges are caused by the down-selection of solar-type dwarfs and sub-giants using Equations~\ref{eq:instab} and~\ref{eq:numax}. As expected, the top-ranked stars are dominated by cool, high-luminosity sub-giants with intrinsically high detection probabilities. Progressing towards lower ranks, a larger number of hot stars appear, a direct consequence of relaxing the $\beta$ amplitude dilution factor described in Section~\ref{sec:rank}. 

The distribution of targets in Figure~\ref{fig:hrd} demonstrates the well known bias of asteroseismic detections against cool, low-mass stars due to their intrinsically low oscillation amplitudes (see, e.g., \citealt{chaplin_predicting_2011}). In total, only six stars ranked among the top \nstars\ in the ATL have luminosities less than solar. This makes the ATL highly complementary to the exoplanet target list, which prioritizes cool dwarfs due to the improved probability of finding small transiting exoplanets. We note that all solar-type stars having a magnitude $T<6$ in the TESS bandpass are automatically included in the TESS 2-minute cadence target list \citep{Stassun2018}, irrespective of their position on the ATL.

\subsection{Expected Yield}

%%%%%%%%%%%%%%%%%%%%%%%%%%%%%%%%%%%%%%%%%%%%%%%%%%%%%%%%%%%%%%%%%%%%%%%%%%%%%%%%%%%%%%%%%%%%

\begin{figure*}
	\centering
	\includegraphics[width=17cm]{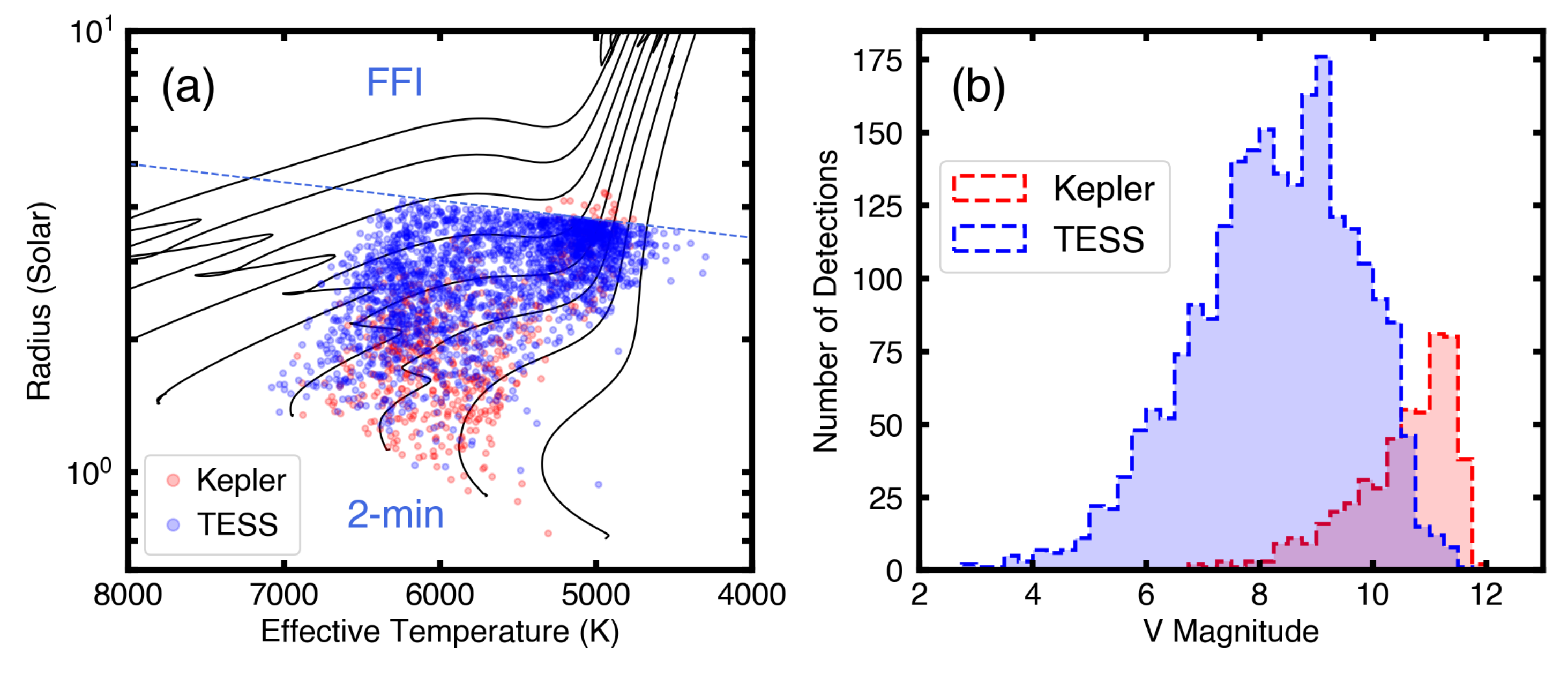}
	\caption{The predicted asteroseismic yield for the first year of TESS science operations (Cycle 1). Panel (a): Radius versus effective temperature for all expected TESS detections (blue) and the detections for dwarfs and sub-giants by \emph{Kepler} (red). The blue dashed line marks the approximate radius limit above which oscillations can be confidently detected using FFI light curves. Black lines show evolutionary tracks. Panel (b): Approximate $V$ magnitude distribution of the expected TESS yield (blue) and the \emph{Kepler} yield (red).}	
	\label{fig:yield}
\end{figure*}

%%%%%%%%%%%%%%%%%%%%%%%%%%%%%%%%%%%%%%%%%%%%%%%%%%%%%%%%%%%%%%%%%%%%%%%%%%%%%%%%%%%%%%%%%%%%

To estimate the expected yield of asteroseismic detections, we performed a Monte-Carlo simulation as follows. For each star in the ATL, we drew a uniform random number $n$ between zero and unity and counted the target as a potential seismic detection if $n < p_{\rm vary}$. We recall here that $p_{\rm vary}$ provides a conservative yield, since the amplitude dilution factor $\beta$ may be overestimated. To determine whether the target would be observed in 2-minute cadence, we adopted a starting ecliptic longitude of zero degrees for the first observing sector and picked the top 450 targets (the per-sector allocation of TASC) in the ATL that fall on silicon in that sector. We then repeated this for each of the 13 sectors in the southern ecliptic hemisphere, adding new detections to the list each time.

The predicted TESS yield for the first full year of science operations, corresponding to Guest Investigator Cycle 1, is $\sim$\,2500 oscillating targets, already a five-fold increase over the yield from the \emph{Kepler} mission. Of these detections, the majority are observed for a single sector ($\sim$\,1500), while $\sim$\,200 targets are expected to be observed for 10 sectors or more. The second year of nominal science operations (Cycle 2) is expected to produce a similar yield, bringing the total expected number of detections to 5000 stars. We emphasize that these estimates only take into account stars on the ATL and ignore potential overlaps with other target lists (such as the CTL and Guest Investigator Program), which would result in a slightly higher yield. They also assume that our adopted noise model provides a good description of the actual, in-flight photometric precision.

Figure~\ref{fig:yield}a compares the predicted asteroseismic yield of TESS to detections for dwarfs and sub-giants from the \emph{Kepler} mission \citep{Chaplin2014}. As expected, the TESS yield is skewed towards evolved sub-giants with intrinsically larger amplitudes, and contains a smaller number of cool dwarfs (for which higher photometric precision is required for a detection). Importantly, Figure~\ref{fig:yield}b demonstrates that the TESS detections will be on average 4-5 magnitudes brighter than \emph{Kepler}, which follows from the difference in aperture size. Similar to the characterization of transiting exoplanets, this will enable significantly more powerful complementary follow-up observations, including measurements of angular diameters using optical long-baseline interferometry \citep[which was only possible for a handful of \emph{Kepler} dwarfs and subgiants][]{huber12,white13}. Overall this demonstrates that TESS will excel in a significantly different parameter space than \emph{Kepler}, in particular for evolved subgiants which exhibit mixed modes that allow powerful constraints on the interior structure \citep[e.g.][]{chaplin_asteroseismology_2013, hekker_giant_2016}.

 %\subsection{Accessing the ATL catalog and codes}
 %\label{sec:access}

% Move to conclusion...

\section{Summary and Conclusions}
\label{sec:sum}

We have presented the construction of the Asteroseismic Target List (ATL) for solar-like oscillators to be observed in 2-minute cadence by the TESS Mission. The main characteristics of the ATL can be summarized as follows:

\begin{itemize}
\item The ATL includes \nstars\ bright main-sequence and subgiant stars that have at least a 5\% probability of detecting solar-like oscillations with TESS. Detection probabilities were calculated from stellar properties estimated from colors, parallaxes and apparent TESS magnitudes. The ranking of targets is based on a mixture of detection probability and the prioritization of hot stars, for which the oscillation amplitudes are poorly understood. 

\item We have validated our derived stellar properties against spectroscopy, asteroseismology and interferometry, finding good agreement. In addition to the asteroseismic detection probabilities, the ATL provides a homogeneous catalog of stellar properties for bright solar-type stars observed by TESS.

\item Based on the nominal TESS photometric performance and the number of target slots assigned to the ATL, we expect that TESS will increase the number of solar-type stars with detected oscillations by an order of magnitude over \emph{Kepler}. Most of the detections will be in evolved subgiants, with only a small number of detections in unevolved main-sequence stars.

\item The \texttt{Python} code used to produce the ATL is publicly available on \texttt{Github}\footnote{\url{https://github.com/MathewSchofield/ATL_public}}\footnote{\url{https://figshare.com/s/aef960a15cbe6961aead}}, allowing full reproducibility of the asteroseismic target selection for comparison with population synthesis models. The ATL itself is available in electronic form\footnote{\url{https://figshare.com/s/e62b08021fba321175d6}}. The columns of the ATL are shown in Table~1.
\end{itemize}

The yield of solar-like oscillators with TESS is expected to continue the asteroseismic revolution initiated by CoRoT and \emph{Kepler}. In particular, TESS is expected to deliver detections in the nearest solar-type stars for which strong complementary constraints (e.g.\ from Hipparcos/Gaia parallaxes and interferometry) are available, allowing powerful inferences on the interior structure of stars and stellar ages, including exoplanet host stars. Our improved understanding of the excitation mechanism of solar-like oscillations probed by the large sample of TESS stars observed in 2-minute cadence will also be helpful to optimize target selection for future missions such as PLATO \citep{rauer14}.

%%%%%%%%%%%%%%%%%%%%%%%%%%%%%%%%%%%%%%%%%%%%%%%%%%%%%%%%%%%%%%%%%%%%%%%%%%%%%%%%%%%%%%%%%%%%

\begin{deluxetable}{ll}
\label{tab:atl}
\tablecaption{Column headers of the ATL}
\tablewidth{0pt}
%\tablehead{
%\colhead{Column}&
%\colhead{Description}}
\tabletypesize{\scriptsize}
\startdata
 & \\
01:& TESS \emph{Input Catalog} (TIC) ID\\
02:& \emph{Tycho}\,2 ID\\
03:& \emph{Hipparcos} ID\\
04:& \emph{Gaia} Data Release 1 ID\\
05:& \emph{Gaia} Data Release 2 ID\\
06:& Maximum number of contiguous observing sectors (1-13)\\
07:& Rank based on $p_{\rm mix}$\\
08:& (Flag) 1: Rank manually adjusted or star added to list afterwards\\
09:& (Flag) 1: High priority star (for 20-sec cadence); 0: 120-sec cadence star\\ 
10:& Ecliptic latitude (deg)\\
11:& Ecliptic Longitude (deg)\\
12:& Galactic latitude (deg)\\
13:& Galactic longitude(deg)\\
14:& Equatorial declination (deg)\\
15:& Equatorial right ascension (deg)\\
16:& TESS-band apparent magnitude (mag)\\
17:& $V$-band apparent magnitude (mag)\\
18:& $I$-band apparent magnitude (mag)\\
19:& Extinction in $I$-band (mag)\\
20:& Extinction in $V$-band (mag)\\
21:& $(B-V)$ color (mag)\\
22:& $(B-V)$ color uncertainty (mag)\\
23:& Reddening of $(B-V)$ color (mag)\\
24:& Parallax (mas)\\
25:& Parallax uncertainty (mas)\\
26:& (Flag) 1: \emph{Hipparcos} parallaxes used; NaN: DR2 parallaxes used\\
27:& Distance (Kpc)\\
28:& Distance uncertainty (Kpc)\\
29:& (Flag) 1: Bailer-Jones et al. (2018) distances are provided\\
30:& Bailer-Jones et al. (2018) distance (upper limit) (pc)\\
31:& Bailer-Jones et al. (2018) distance (median value) (pc)\\
32:& Bailer-Jones et al. (2018) distance (lower limit) (pc)\\
33:& Luminosity $L$ (in $L_{\odot}$)\\
34:& $\nu_{\rm max}$ ($\rm \mu Hz$)\\
35:& Radius $R$ (in $R_{\odot}$)\\
36:& $T_{\rm eff}$ (K)\\
37:& Global asteroseismic SNR ($\beta=1$)\\
38:& Global asteroseismic SNR ($0 \le \beta \le 1$)\\
39:& $p_{\rm mix}$ composite probability\\
40:& $p_{\rm vary}$ probability ($0 \le \beta \le 1$)\\
41:& $p_{\rm fix}$ probability ($\beta = 1$)\\
\enddata
\end{deluxetable}

%%%%%%%%%%%%%%%%%%%%%%%%%%%%%%%%%%%%%%%%%%%%%%%%%%%%%%%%%%%%%%%%%%%%%%%%%%%%%%%%%%%%%%%%%%%%

\acknowledgments

M.S. acknowledges support from the University of Birmingham. W.J.C., G.R.D., A.M. and W.H.B. acknowledge support from the UK Science and Technology Facilities Council (STFC). D.H. acknowledges support by the National Aeronautics and Space Administration (NASA) under grant NNX14AB92G issued through the \emph{Kepler} Participating Scientist Program. T.L.C. acknowledges support from the European Union's Horizon 2020 research and innovation programme under the Marie Sk\l{}odowska-Curie grant agreement No.~792848 and from grant CIAAUP-12/2018-BPD. T.S.M. acknowledges support from NASA grant NNX16AB97G. A.S. acknowledges partial support from grants ESP2017-82674-R (Spanish Ministry of Economy) and SGR2017-1131 (Generalitat de Catalunya). Funding for the Stellar Astrophysics Center is provided by The Danish National Research Foundation (Grant agreement no.: DNRF106). \textbf{Finally, we thank the anonymous referee for helpful comments on the manuscript.}

\bibliography{ATL}

\end{document}